\documentclass[a4paper, pre, twocolumn,superscriptaddress,floatfix,amssymb,showpacs]{revtex4-1}\usepackage{enumerate}
\usepackage{times}
\usepackage{graphicx}
\usepackage{amsmath}
\usepackage[usenames]{xcolor}
\usepackage{natbib}
\bibpunct{[}{]}{,}{n}{}{;}

\usepackage[pdfpagemode=None,colorlinks=true,urlcolor=black,%
linkcolor=blue,citecolor=blue,pdfstartview=FitH]{hyperref}
\bibpunct{[}{]}{,}{n}{}{;}

\graphicspath{{fig/}} 
\definecolor{plotbrown}{RGB}{180,4,38}
\definecolor{plotpurple}{RGB}{128,0,128}

\begin{document}

\title{Polymer ejection from strong spherical confinement}

\author{J. Piili}
\author{R. P. Linna}
\email{Corresponding author: riku.linna@aalto.fi}
\affiliation{Department of Computer Science, Aalto University, P.O. Box 15400, FI-00076 Aalto, Finland}

\begin{abstract}
We examine the ejection of an initially strongly confined flexible polymer from a spherical capsid through a nanoscale pore. We use molecular dynamics for unprecedentedly high initial monomer densities. We show that the time for an individual monomer to eject grows exponentially with the number of ejected monomers. By measurements of the force at the pore we show this dependence to be a consequence of the excess free energy of the polymer due to confinement growing exponentially with the number of monomers initially inside the capsid. This growth relates closely to the divergence of mixing energy in the Flory-Huggins theory at large concentration. We show that the pressure inside the capsid driving the ejection dominates the process that is characterized by the ejection time growing linearly with the lengths of different polymers. Waiting time profiles would indicate that the superlinear dependence obtained for polymers  amenable to computer simulations results from a finite-size effect due to the final retraction of polymers' tails from capsids.
\end{abstract}


\maketitle

\section{Introduction}
Processes involving macromolecules in confinements are studied intensely due to their significance in biology and potential technological and medical applications. An important class of such processes is the capsid ejection where a polymer is initially in a compact conformation inside a capsid and then ejects outside through a pore a few nanometers wide. By far the most important biological process of this class is the viral packaging in and ejection from bacteriophages~\cite{muthukumar1, muthukumar2, smith,grayson, ali, ghosal, cacciuto2, sakaue_polymer_decompression, riku_dynamics_of_ejection}. The ejection of double-stranded (ds) DNA is clearly the most studied case. Ejection processes of biopolymers such as RNA and single-stranded (ss) DNA, although less studied, are found {\it in vivo} and are biologically highly relevant. The recent technological advancement in using engineered viral capsids for drug delivery~\cite{glasgow} further emphasizes the need to understand the fundamentals of biopolymer packaging in and ejection from a capsid. While dsDNA is modeled as a semiflexible polymer, ssDNA and RNA among many other biopolymers are fully flexible~\cite{alberts}.

There is a wide gap between the conditions under which biopolymers eject from viral capsids found {\it in vivo} and those taken to prevail in computer simulations and theoretical analysis. In the bacteriophages the viral DNAs and RNAs are packed to almost crystalline densities~\cite{smith}, realization of which is beyond most computer simulations. Computational investigations of some specific characteristics using close-to-realistic model polymers packed to high densities are typically done using some form of probabilistic Metropolis sampling, see {\it e.g.}~\cite{marenduzzo_topological_dna_ejection}. On the other hand investigations using dynamically more realistic molecular dynamics (MD) based methods typically aim at detailed modeling of a specific polymer and are not very conclusive with regard to general characterization of the polymer ejection. Due to restrictions placed by the detailed model on polymer lengths and on statistics no attempt is made to relate to the existing theoretical formulations.

Our motivation is to determine how applicable the available theoretical formulations to the real-world ejection processes are. The available theories do not explicitly take into account the bending rigidity in the case of dsDNA. Still, they are commonly used as a basis for understanding also the ejection of dsDNA. The current understanding is that a packed dsDNA assumes a spool-like conformation inside the capsid. However, in the theoretical formulations polymers start from disordered conformations. Blob-scaling picture that describes polymers under semi-dilute conditions~\cite{degennes} was used in the derivations. This, in effect, means that the theories are not guaranteed to be valid for very strongly confined polymers found {\it in vivo}.

Here, we set out to study the ejection of fully flexible polymers from spherical confinements. The studied confinements are clearly stronger than those in previous studies but still much weaker than those found {\it in vivo}. The flexible freely-jointed chains (FJC) start from random initial conformation, so that close comparison with the theoretical predictions can be made. In short, we use a generic polymer model but realistic MD based dynamics to characterize the ejection dynamics of fully flexible polymers initially under very strong spherical confinement. The effect of polymer bending rigidity and the accompanying spooled initial conformation that are important in the case of dsDNA will be reported in a future publication.

Due to the computationally effective implementation of the capsid geometry sufficient statistics can be obtained for polymer ejection processes starting from the high initial monomer densities. In our capsid model boundary conditions are imposed instead of introducing a repulsive potential between polymer and the capsid. This way external forces that could affect the ejection dynamics are avoided. Accordingly, we can obtain the polymer's excess energy due to confinement. We show that the form of this energy, which decidedly deviates from the predictions based on the blob-scaling picture widely used for characterization of confined polymers, determines the ejection dynamics.

We model Brownian heath bath with stochastic rotation dynamics (SRD) where also hydrodynamics can be included. In order to keep the analysis simple, we use the method here with hydrodynamical interactions switched off. The effects of hydrodynamics will be analyzed in detail in the forthcoming paper. In what follows, we first outline the understanding previously obtained from the blob-scaling picture. We then describe our simulation method after which the results are presented and analyzed. Lastly, we summarize and recap the main conclusions.

\section{Current theoretical understanding}

The evolution of the theoretical understanding of the ejection of a fully flexible polymer from a capsid was initiated by an investigation by Muthukumar where Monte Carlo simulations were used~\cite{muthukumar1}. Assuming the excess energy due to confinement to be $\Delta F/k_BT \sim N_0/R_0^{1/\nu}$ a scaling prediction for the ejection time of the form $\tau \sim N_0 R_0^{1/\nu} = N_0(N_0/\rho_0)^{1/{3\nu}}$ was obtained, where $N_0$ is the degree of polymerization, $R_0$ the capsid radius, $\rho_0$ the initial monomer density in the capsid, and $\nu$ the Flory exponent. This scaling was confirmed by MC simulations. In a later work~\cite{cacciuto2} the excess energy due to spherical confinement was taken to follow the scaling $\Delta F/k_BT \sim (R_0/R_G)^{3/(3\nu-1)} \sim N_0 \phi^{1/(3\nu-1)}$, where $\phi_0 = N_0 a^3/R_0^3$ is the initial monomer volume fraction and $a$ is the monomer length. This scaling law, first introduced by Grosberg and Khokhlov~\cite{grosberg}, was shown to be correct for the spherically confined polymer at $\phi_0$ where the blob scaling in the semidilute conditions is valid~\cite{sakaue_confined}. This led to the scaling relation $\tau \sim N_0^{1+\nu}\phi_0^{1/(1-3\nu)}$, again confirmed by MC simulations. The different regimes and short ranges explain why different scaling relations in these studies were corroborated by simulations.

The  unified framework~\cite{sakaue_polymer_decompression} for polymer decompression processes presented a fairly complete view of the ejection process. Assuming uniform polymer conformation the excess confinement energy inside a spherical capsid was derived in the same form as in~\cite{grosberg},
\begin{equation}
\Delta F/k_BT \approx \Big(\frac{a}{R_0}\Big)^{3/(3\nu-1)}N(t)^{3\nu/(3\nu-1)},
\label{conf_en}
\end{equation}
where $N(t)$ is the number of monomers inside the capsid.

The resultant driving force $\sim k_BT/\xi(t)$ was taken to be exerted on the monomer residing at the pore. The  overall dissipation takes place close to the pore, within the range of the correlation length $\xi(t)$, where there exists a velocity gradient of segments of the order $\sim a \dot{N}(t)/\xi(t)$. Accordingly, the dissipation term was evaluated as $T\dot{S}(t) = \eta[\dot{N}(t)/\xi(t)]^2\xi(t)^3$. The excess confinement energy is dissipated at the rate $\Delta \dot{F}(t) = -T\dot{S}(t)$, from which the time evolution was obtained as 
\begin{equation}
N(t) = N_0(1+t/\tau_1)^\beta, 
\label{sakaue_nt}
\end{equation}
where the exponent $\beta = (1-3\nu)/[2(1-\nu)]$ and the time constant $\tau_1 \simeq \tau_0 \phi_0^{(1+\nu)/(1-3\nu)}N_0$. $\tau_0 \simeq \eta a^3/(k_BT)$ is the monomer scale time constant, where $\eta$ is the solution viscosity. For the pressure-driven part the scaling $\tau \sim \tau_0 \phi_0^{-(\nu +2)/(3\nu)}N_0^{(2+\nu)/(3\nu)}$ was obtained.~\cite{sakaue_polymer_decompression}

Blob-scaling assumption was used in the theories outlined above. Also in the strong confinement, as defined in~\cite{sakaue_polymer_decompression,sakaue_confined}, the confined polymer chain is assumed reminiscent of a semidilute polymer solution and the description is based on the blob-scaling assumption.

\section{The computational model}
\begin{figure}
 \includegraphics[width=\linewidth]{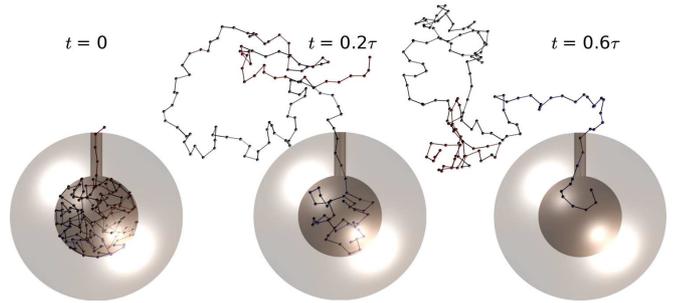}
\caption{(Color online) Snapshots of a simulated polymer ejection. (Images created using VMD~\cite{vmd} and POV-Ray~\cite{povray})}
\label{fig:capsidPicture}
\end{figure}

For the simulation of the ejection dynamics we use a hybrid computational method where the time-integration of the polymer is performed by MD implemented by the velocity Verlet (vV) algorithm~\cite{swope_velocity_verlet, frenkel_moldy}. The polymer is immersed in a solvent that is modeled by SRD~\cite{malevanets_orig, malevanets_mesoscopic}. Here, we exclude the hydrodynamics for the better understanding of the fundamental ejection dynamics and more straightforward comparison with the existing understanding based on the blob picture. The spherical capsid is modeled as a shell with rigid walls imposing slip and no-slip boundary conditions for polymers and solvent, respectively. The pore is modeled as a cylindrical hole in the shell. The simulation geometry and snapshots of an ejecting polymer are depicted in Fig.~\ref{fig:capsidPicture}. The polymer bond length at rest is approximately $1$. The pore radius $R_p$ is $0.4$ for the polymer and $0.8$ for the solvent. The thickness of the capsid wall is $3$. The capsid geometry was created using computationally effective constructive solid geometry technique~\cite{wyvill_csg}.

The polymer is modeled as a chain of point-like beads with mass $m_b$. Adjacent beads are connected via the FENE potential $U_F = - \frac{K}{2} r_{\rm max}^2 \ln{\left( 1 - \left(\frac{r}{r_{\rm max}} \right)^2 \right)}\;,\; r < r_{\rm max},$ where $r$ is the distance between adjacent beads and $K$ and $r_{\rm max}$ are potential parameters describing the strength and maximum distance limit of adjacent beads. A Lennard-Jones potential acts between all beads: $U_{LJ} = 4.8 \epsilon \left[\left(\frac{\sigma}{r_{ij}}\right)^{12} - \left(\frac{\sigma}{r_{ij}}\right)^6\ \right] + 1.2\epsilon$ for $r_{ij} \leq \sqrt[6]{2}\sigma$ and $U_{LJ} = 0$ for $r_{ij} > \sqrt[6]{2}\sigma$. $\epsilon$ and $\sigma$ are potential parameters and $r_{ij}$ is the distance between beads $i$ and $j$. The potential parameters are chosen as $\sigma = 1.0$, $\epsilon = 1.0$, $K = 30/\sigma^2$, and $r_{\rm max} = 1.5\sigma$ in reduced units~\cite{allen}.

We use initial monomer density $\rho_0 = N_0/(\frac43 \pi R_0^3)$ instead of volume fraction $\phi_0$. Accordingly, the number of monomers $N_0$ corresponding to $\rho_0 = 1$ is by the factor $\frac{4}{3} \pi$ larger than $N_0$ corresponding to $\phi_0 = 1$. On the other hand, $\phi_0$ was used for hard spheres, the use of which is not possible in high-density MD simulations. A value of $\phi_0$ constitutes a slightly lower compression in a system using soft potentials than in one using hard-sphere potentials. This effect is more than compensated by our using $\rho_0$ instead of $\phi_0$, so effectively the systems simulated here start from more compressed states than those dealt with in~\cite{muthukumar1,cacciuto2,sakaue_polymer_decompression}. In the initial conformations four beads are inside the pore so the total length of the polymers is $N_0+4$.

SRD solvent consists of point-like particles whose dynamics can be divided into two steps. In the streaming step the solvent particle positions are ballistically propagated in discrete time steps. The interactions between particles are taken into account in the collision step. Here the random parts of the velocities for the polymer and solvent particles divided into cubic cells of unit edge lengths are rotated by the angle $\alpha = 3\pi/4$ around an axis chosen randomly for each cell. In the present case of Brownian heat bath velocities are randomly exchanged between all particles after the collision step. The solvent is kept at the constant temperature of $k_B T = 1.0$ by scaling the random parts of particle velocities such that the equipartition theorem holds at all times. In order to maintain Galilean invariance, the grid is shifted randomly at each time step~\cite{ihle_galilean}. For our simulations the vV time step $\delta t = 0.0002$ and the SRD time step $\Delta t = 0.5$. MD and SRD steps are performed in turns such that after $\Delta t/\delta t$ vV steps a single SRD step is performed. More detailed accounts of the SRD method can be found in~\cite{malevanets_orig, malevanets_mesoscopic}.

The unit of length in simulations is typically mapped to real-world length scale in various ways depending {\it e.g.} on the polymer that is modeled. Here we present a mapping when the FJC polymer of our simulation is taken to model a single stranded DNA (ssDNA). The persistence length of ssDNA is of the order of $4$ nm~\cite{tinland}. In our FJC the the persistence length is $\lambda_p = \frac12b$, where $b$ is the bond length of about $1$ in our simulation units. Accordingly, one simulation unit $\sigma$ corresponds to about $8$ nm and a polymer of length $N_0=200$ to a ssDNA approximately $1600$ nm long. Since the base in ssDNA is about $0.37$ nm long~\cite{rechendorff_length_per_bead}, a polymer of length $N_0=200$ corresponds to about $4324$ bases. In our simulations, the capsid inner radius varies from 1.6 ($N = 25, \rho_0 = 1.5$) to 5.1 ($N_0 = 283, \rho_0 = 0.5$), which corresponds to the range from $12.8$ to $40.8$ nm.

To gain understanding about the time scales involved in the ejection process, we measured the relaxation times of free polymers starting from fully stretched conformations. The measured radii of gyration of polymers follow approximately
\begin{align}
R_g(t) \sim \left[ R_g(0) - R_g(\infty)\right]\exp{(-t/\tau_{\rm relax})}+R_g(\infty) ,
\end{align}
where relaxation times $\tau_{\rm relax} = 1642,$ $8883,$ and $28780$ for $N_0=50, 100,$ and $200$, respectively. Hence, in our simulations the ejection takes from $4.2$ to $12.5$ times the relaxation time for $N_0=50$ and from $2.5$ to $7.1$ times the relaxation time for $N_0=200$. In principle, this can be compared to relaxation time of a real ssDNA. In Ref.~\cite{liu_relaxation_times} the authors measured relaxation times for single stranded $\lambda$-DNA. They find that the relaxation time is $\tau_{R,\lambda} = 0.19$ s in solution of viscosity 15 cP. The relaxation times from experiments and our simulations are by no means comparable. In experiments polymer conformations evolve by Zimm dynamics, of course, whereas our simulated polymers perform Rouse dynamics. In addition, the length of the  $\lambda$-DNA is $48 502$ bases long~\cite{gao_lambda_dna}, which is almost ten time as long as in our simulations. Also the viscosity in their experiments is much larger than in our simulations.

\section{Results}

\subsection{Ejection time and waiting times}

\begin{figure}[!ht]
\includegraphics[width=.49\linewidth]{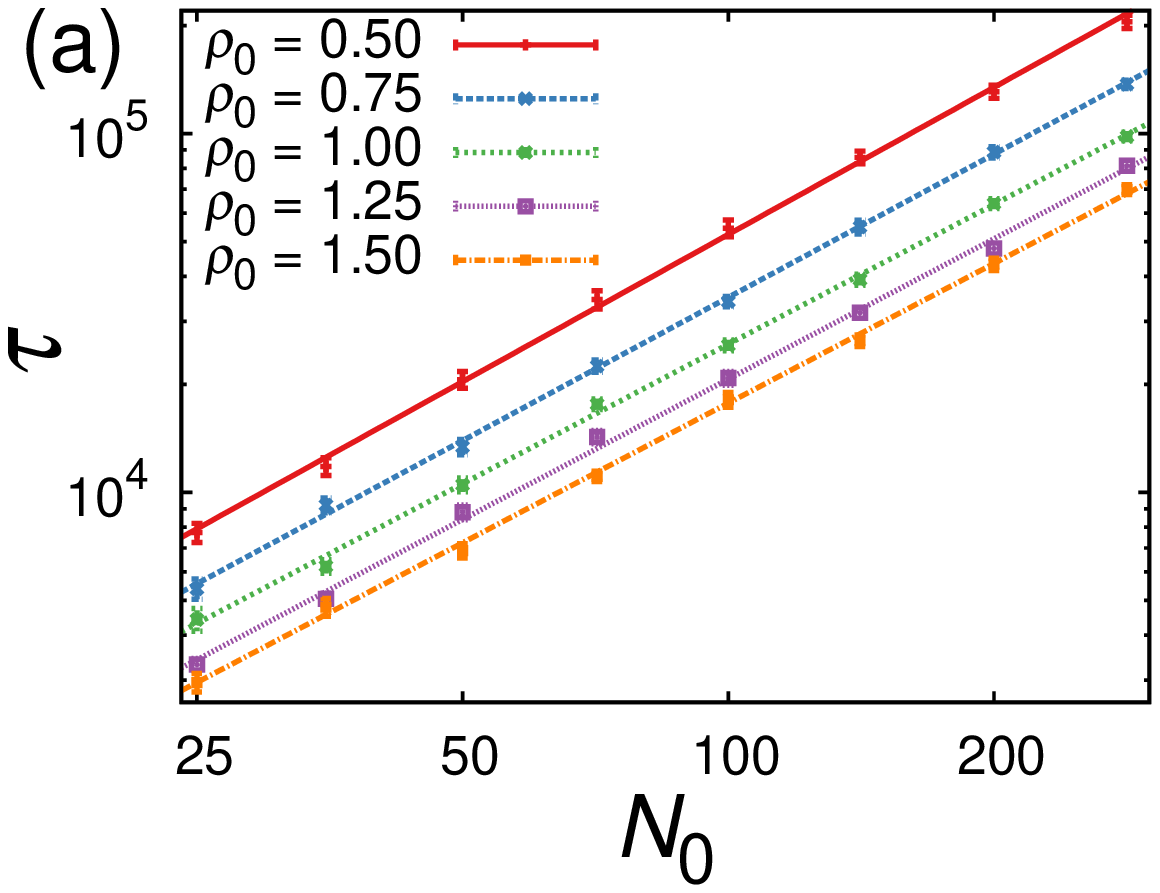}
\includegraphics[width=.49\linewidth]{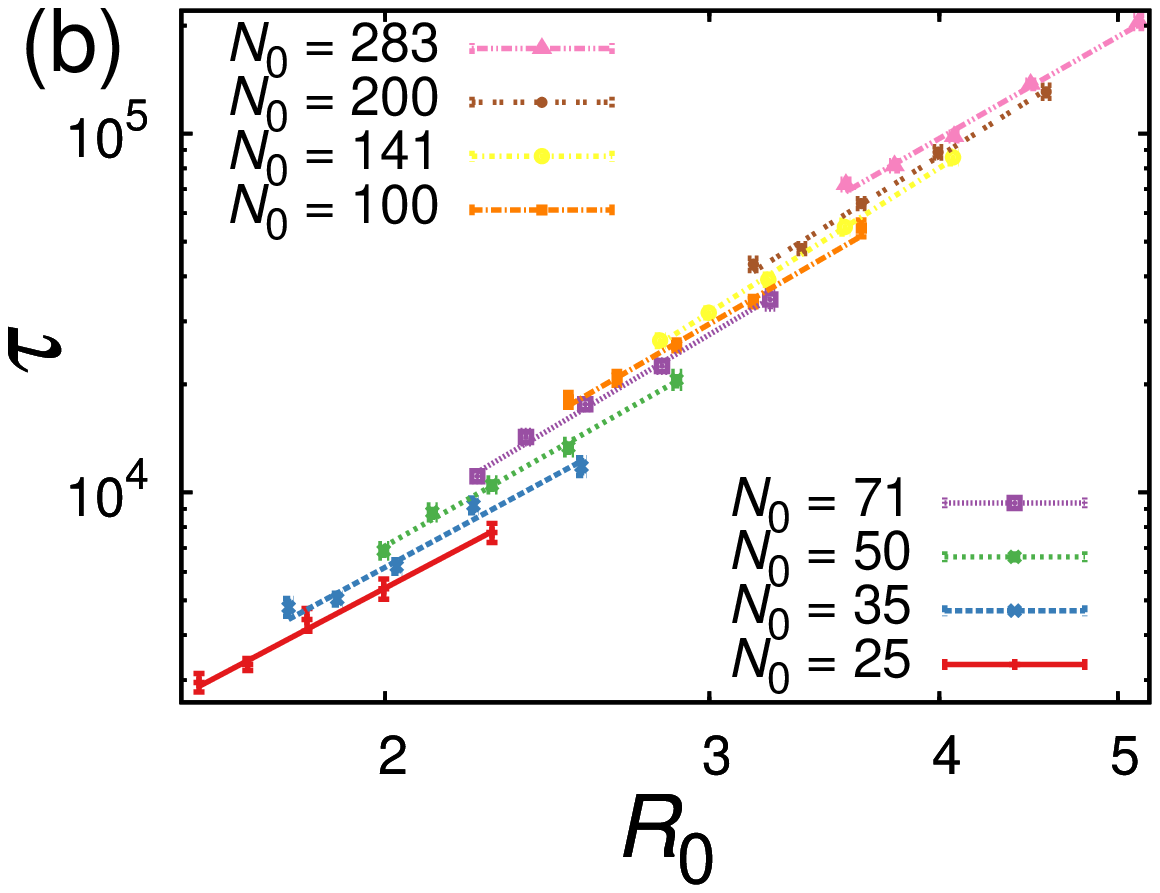}
\caption{(Color online) (a) Ejection time $\tau$ as a function of polymer length $N_0$ for different initial monomer densities $\rho_0$. The points are averages over 40-50 runs. The lines are the fitted curves of the form $\tau \sim  N_0^\beta$. $\beta = 1.362 \pm 0.05$, $1.325 \pm 0.04$, $1.293 \pm 0.04$, $1.299 \pm 0.03$, and $1.300 \pm 0.04$ for $\rho_0 = 0.5$, $0.75$, $1$, $1.25$, and $1.5$, respectively. (b) $\tau$ as a function of capsid radius $R_0$ for different $N_0$, $\tau \sim R_0^\zeta$. From bottom to top: $\zeta = 2.71$, $2.77$, $2.90$, $3.05$, $2.97$, $3.23$, $3.19$, and $2.92$ for $N_0 = 25$, $35$, $50$, $71$, $100$, $141$, $200$, and $283$, respectively. Both figures in log-log scale.} 
\label{fig:tauVsN}
\end{figure}

Fig.~\ref{fig:tauVsN}~(a) shows the ejection time $\tau$ vs number of monomers $N_0$ initially in random conformations inside the capsid for different initial densities $\rho_0$. In accordance with previous findings~\cite{muthukumar1,muthukumar2,smith,grayson,ghosal,cacciuto2,sakaue_polymer_decompression,riku_dynamics_of_ejection}, scaling $\tau \sim N_0^\beta$ is obtained. Here, $\beta = 1.362 \pm 0.05$, $1.325 \pm 0.04$, $1.293 \pm 0.04$, $1.299 \pm 0.03$, and $1.300 \pm 0.04$ for $\rho_0 = 0.5$, $0.75$, $1$, $1.25$, and $1.5$, respectively. (As stated in the Introduction, using initial random conformations is essential for evaluating the validity of the blob-scaling picture.  We also checked that for FJC polymers the initial conformation does not have a strong influence on the ejection dynamics. For the simulated densities $\rho_0 \in \{0.5, 1.0, 1.5\}$ we obtained essentially the same $\beta$ when the ejections started from spooled conformations.) $\beta$ decreases slightly with increasing initial monomer density for $\rho_0 \le 1$ in accordance with our previous findings~\cite{riku_dynamics_of_ejection} and statement on the case of driven translocation when the pore friction increases~\cite{lehtola1,lehtola2}. Most of the effective friction $\gamma$ is exerted on the ejecting polymer in the vicinity of the pore through which the pressure pushes it. $\gamma$ increases with $\rho_0$.  As $\gamma$ increases  $\tau \sim N_0^\beta$, $\beta \to 1$. In accordance with this reasoning, for polymers of $N_0 \ge 50$, $\tau \propto 1/\rho_0$ is obtained very precisely.

Fig.~\ref{fig:tauVsN}~(b) shows $\tau$ as a function of capsid radius $R_0$. Comparing the obtained scaling $\tau \sim R_0^\zeta$, where $\zeta \in [2.71,\ 3.23]$, with the scaling in Fig.~3~(a) of~\cite{sakaue_polymer_decompression}, where $\zeta$ decreases from $5$ toward $2$ when $R_0/a$ decreases from $10^2$ toward $1$, indicates that the ejection takes place in the very strongly confined regime, that is to say, in stronger confinement than required for the strong confinement as defined in~\cite{sakaue_polymer_decompression,sakaue_confined}. We also measured the radii of gyration $R_g$ for the polymers' initial conformations inside the capsids. Spherical scaling $R_g \sim N_0^{\theta}$, where $\theta \approx 0.33$, was obtained for all $\rho_0$, also confirming strong confinement. In spite of the very strong confinement the ejected polymer segment remains close to equilibrium, since we measured $R_g(s) \sim s^{0.6}$ for the ejected segment at different stages. Magnitudes for $R_g(s)$ were also close to the equilibrium values.

\begin{figure}[!ht]
\includegraphics[width=.49\linewidth]{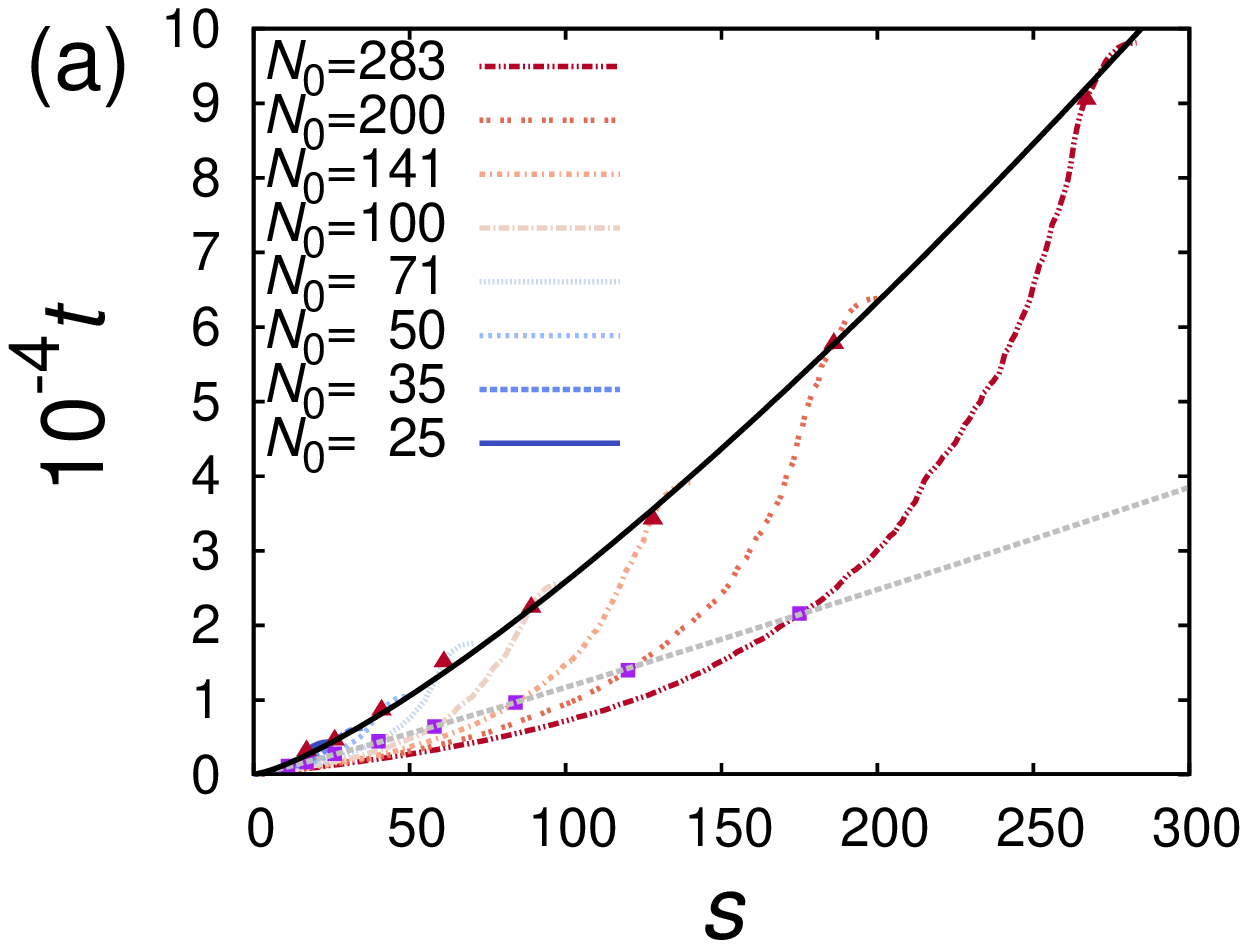}
\includegraphics[width=.49\linewidth]{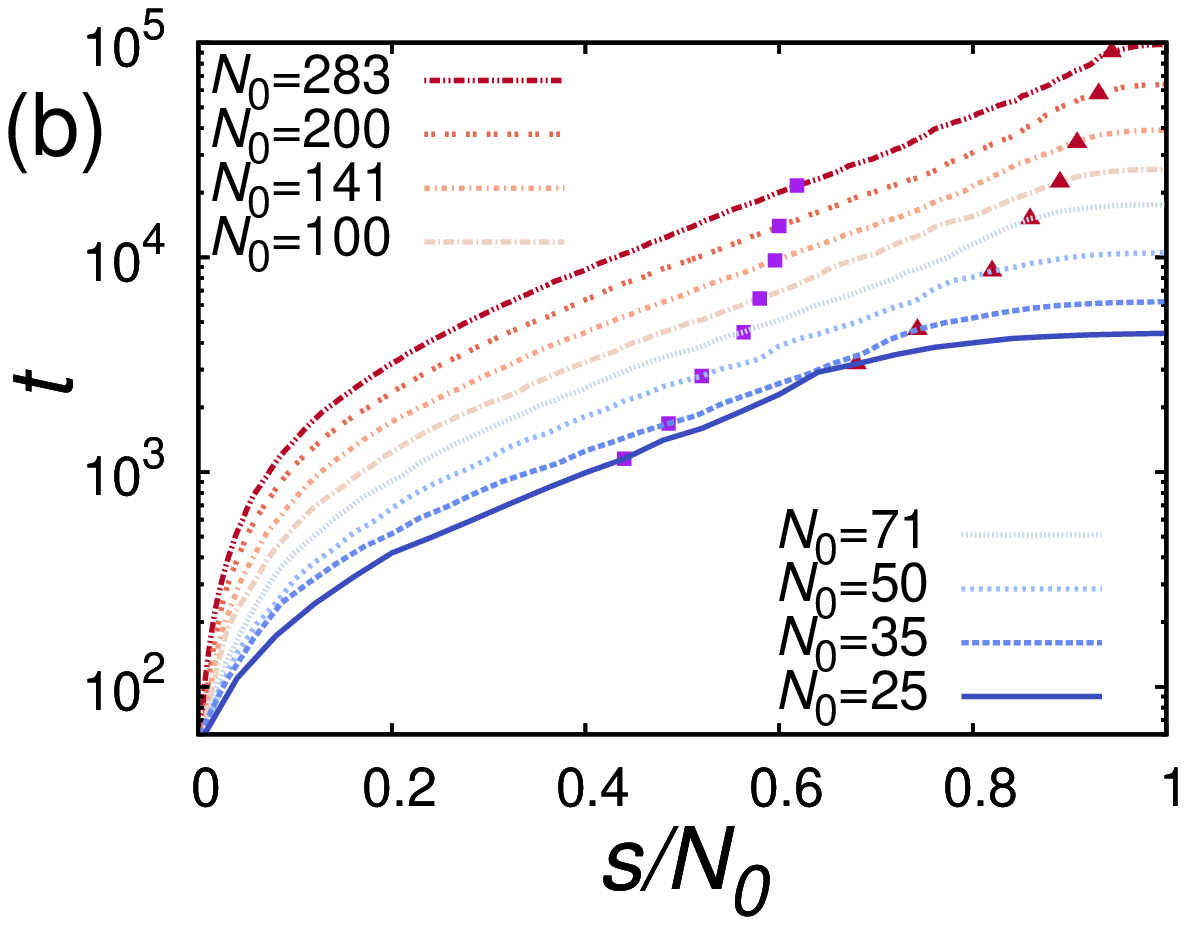}
\includegraphics[width=.49\linewidth]{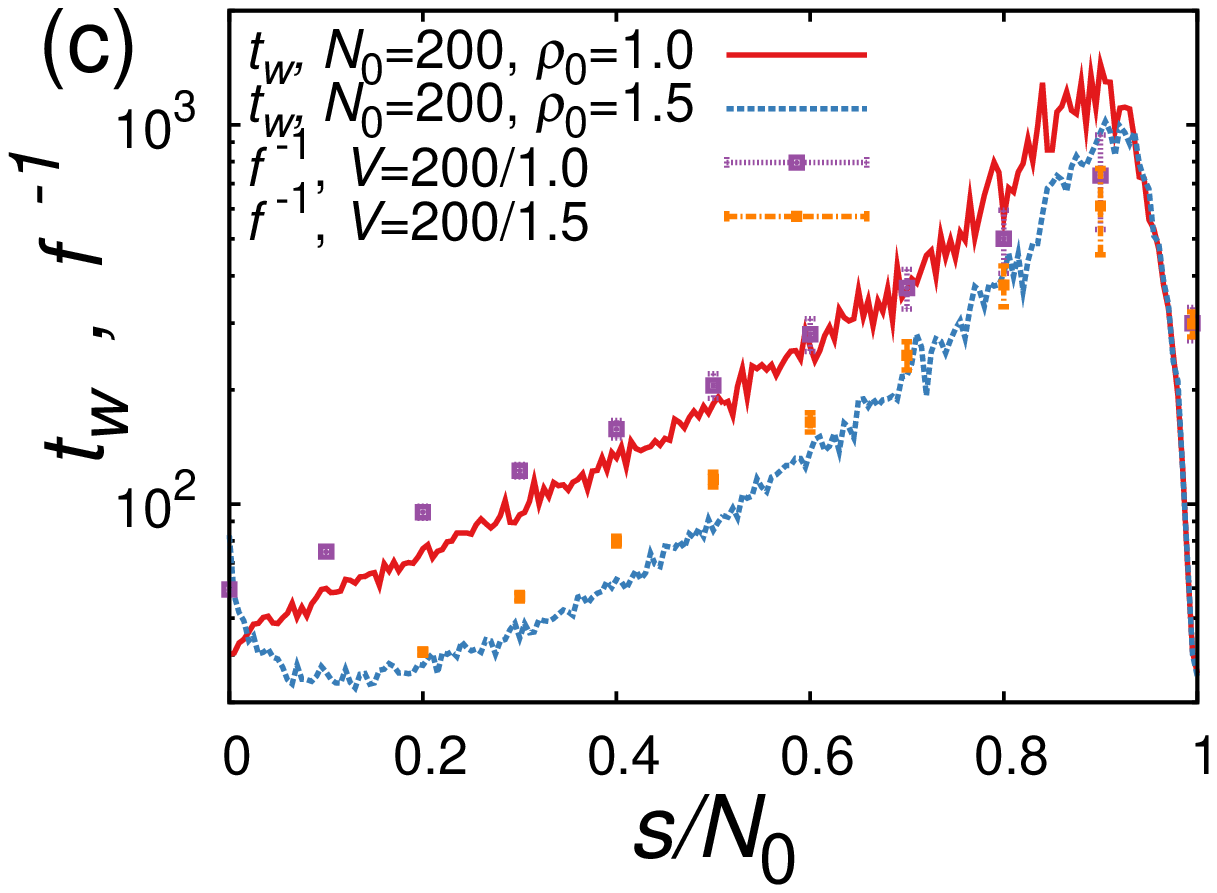}
\includegraphics[width=.49\linewidth]{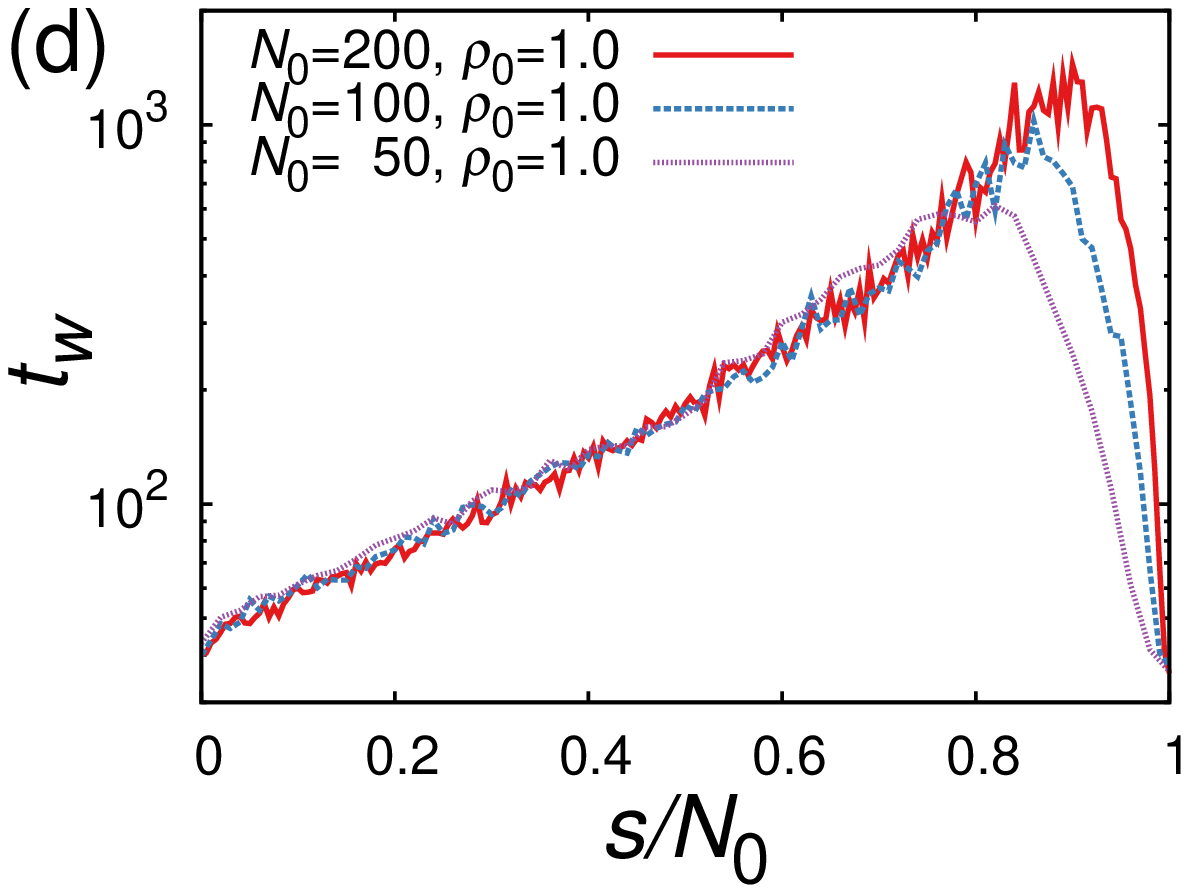}
\caption{(Color online) (a) The cumulative waiting time of simulations where $\rho_0=1.0$. Averages over 50 runs. The times when the end monomers of polymers of different lengths escape scale as $\tau \sim N_0^{1.29}$ (solid line). Reaction coordinates at instants when the inner part of the polymer exerts no force to the bead at the pore (\textcolor{plotpurple}{$\blacksquare$}): $s(\tau_2^*) \sim s^{1.0866}$ (dotted gray line). Reaction coordinate when number of beads inside the capsid $N = g_0$ (\textcolor{plotbrown}{$\blacktriangle$}) (b) The cumulative waiting time vs normalized reaction coordinate $s/N_0$ for $\rho_0 = 1.0$ and different $N_0$. Squares (\textcolor{plotpurple}{$\blacksquare$}) and triangles (\textcolor{plotbrown}{$\blacktriangle$}) correspond to those in (a). (c) Waiting times $t_w$ and the inverse of the measured force at the pore $1/f$ (arbitrary units) for $\rho_0 = 1.0$ and $\rho_0 = 1.5$ when $N_0 = 200$. $V$ is the capsid volume. (d) Waiting times $t_w$ vs. normalized reaction coordinate $s/N_0$ for $\rho_0 = 1.0$ and $N_0 = 50, 100,$ and $200$.}
\label{fig:waitingTime}
\end{figure}

Although ejection times $\tau$ scale with $N_0$, as seen from the endpoints of the cumulative waiting times $t(N_0)$, cumulative waiting times $t(s)$ as such do not scale but grow exponentially with $s$, see Figs.~\ref{fig:waitingTime}~(a) and (b). Accordingly, waiting times of the ejecting monomers $t_w(s) = t(s) - t(s-1)$ follow the exponential form, $t_w \sim \exp(\kappa s)$, as seen in Fig.~\ref{fig:waitingTime}~(c), in accord with the waiting time profile found in~\cite{riku_dynamics_of_ejection} for the symmetric pore.

\begin{figure}
\includegraphics[width=.47\linewidth]{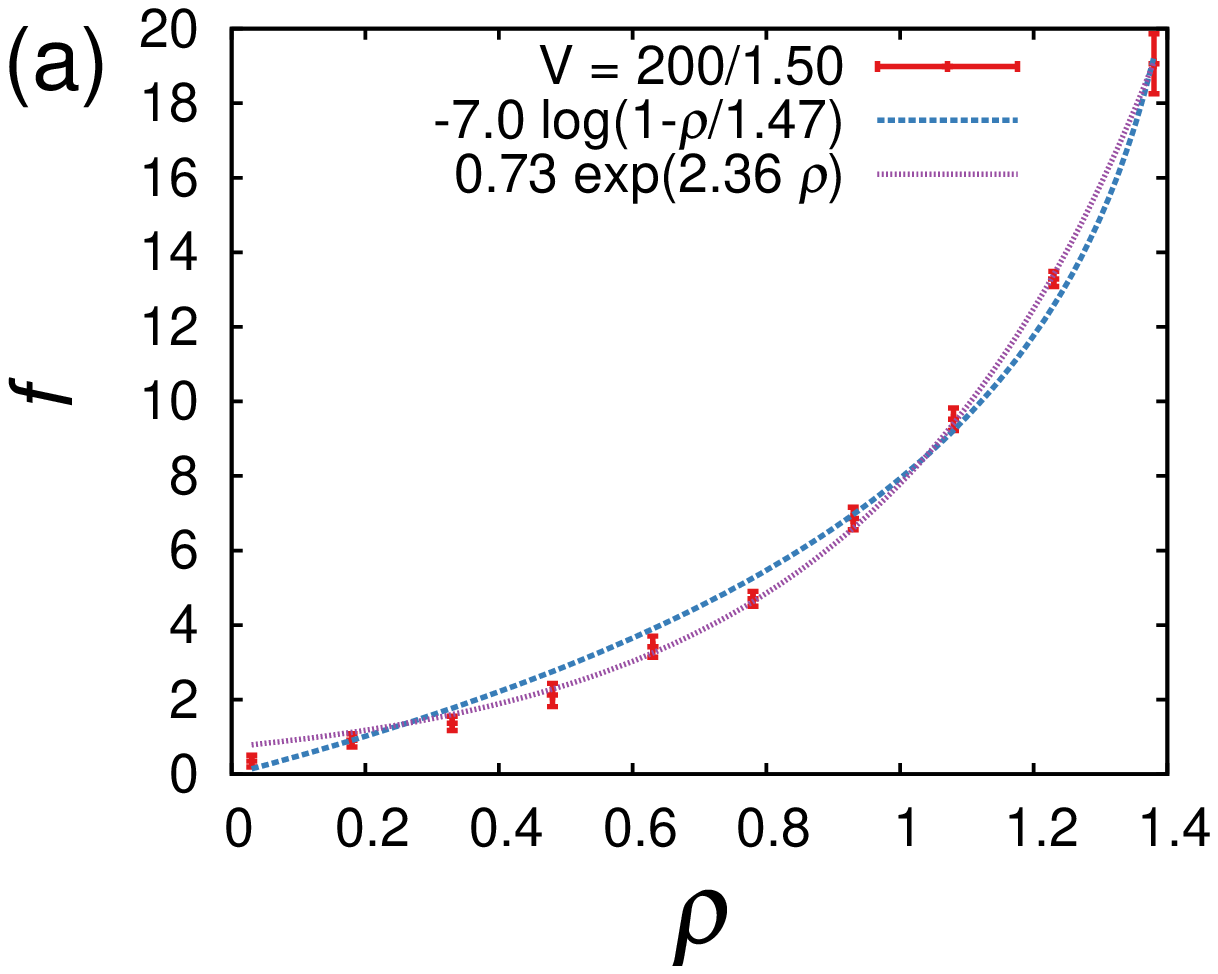}
\includegraphics[width=.505\linewidth]{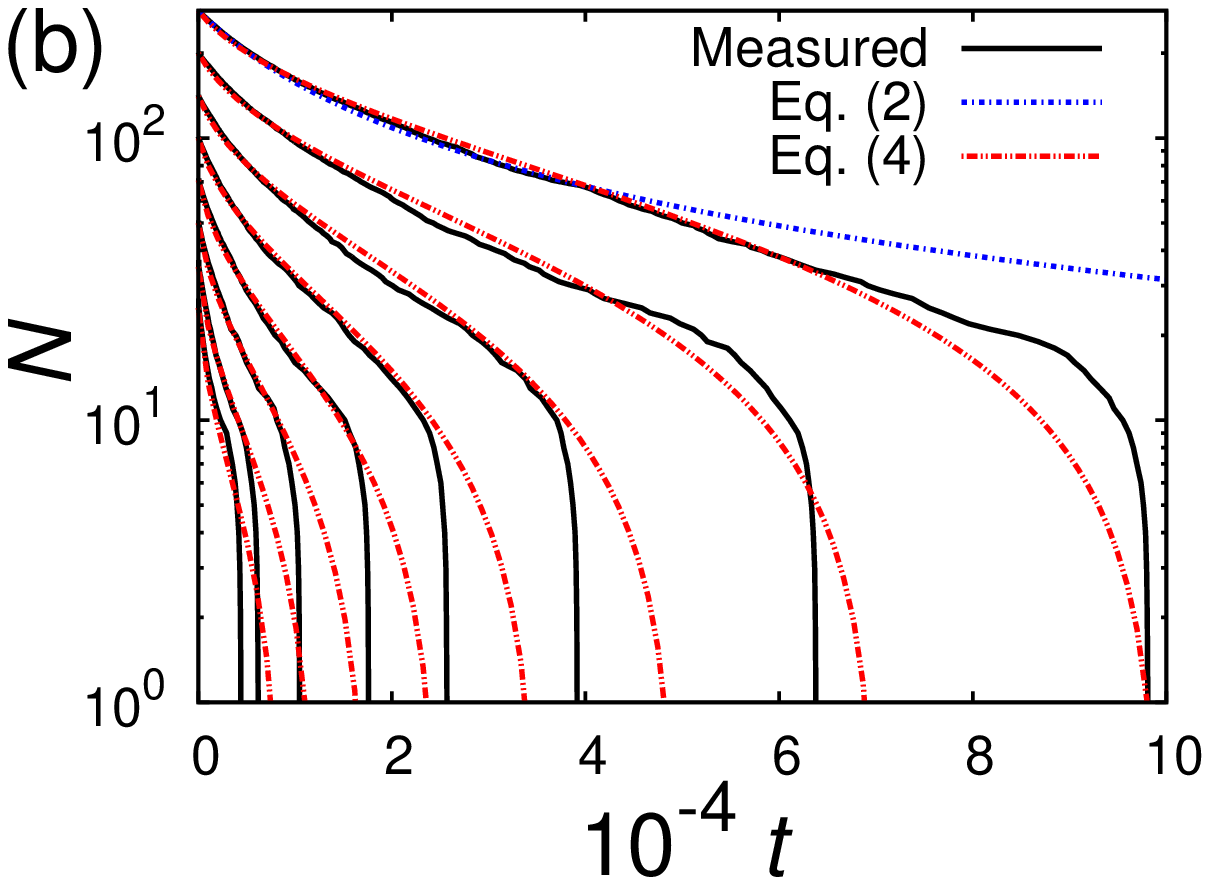}
\caption{(Color online) (a) Force at the pore entrance in equilibrium as a function of the monomer density inside the capsid $\rho_0$ for a capsid of volume $V = 200/1.50$ ($N_0 = 200$ and $\rho_0=1.50$). The curve that is higher in the range $\rho \in (0.3, 1.1)$ is the best fit to the data of the force predicted by the Flory-Huggins theory. The other curve is an exponential curve fitted to the data. (b) Number of beads inside the capsid $N(t)$ during the ejection for $N_0=25,35,50,71,100,141,200,$ and $283$ with fits to Eq.~\eqref{Ntime} and the largest $N_0$ fitted to Eq.~\eqref{sakaue_nt}.}
\label{fig:forceVsDensity}
\end{figure}

\subsection{Excess energy due to confinement}

The form $t_w \sim \exp(\kappa s)$, or equivalently, $t_w(N(t)) \sim \exp(-\kappa N(t))$ is at odds with $\Delta F$ based on the blob-scaling picture, Eq.~(\ref{conf_en}). To determine how largely $\Delta F$ determines $t_w(s)$ we measured the force $f$ that has to be exerted on the monomer at the pore to keep it in position at different $s$. Fig.~\ref{fig:waitingTime}~(c) shows $t_w$ and $1/f$ for $\rho_0 = 1.0$ and $1.5$. $1/f \sim t_w$ is seen to hold well for longer chains throughout the ejection until the start of the final retraction of the remaining segment in the capsid. For $\rho_0 = 1.5$ this relation holds better, as is expected due to increased pressure driving the ejection. The initial deviation is caused by increased jamming at higher densities. Hence, ejection rate is dominantly determined by $\Delta F(N)$ at all stages except for the final retraction.

What is the origin of the exponential dependence $f = C \exp[B N(t)]$? Generally, it can be stated that it arises due to excluded volume interactions as inside the capsid monomers interact individually rather than as ensembles of blobs and the higher order interaction terms become important. Cacciuto and Luijten point out that the blob scaling description breaks down at sufficiently high concentration~\cite{cacciuto_nano}. They proposed a different scaling to set in at this concentration due to screening effects.

Indeed, the average monomer number per blob is very low at concentrations used in our simulations. The blob radius $r_b = A_0 N_b^\nu$, where $N_b$ is the average number of monomers in a blob. $A_0 = 0.6$ was used for the self-avoiding chains in~\cite{cacciuto_nano}. The average density within a blob $\rho_b = N_b/(4/3 \pi r_b^3)$ has to equal the density inside the capsid, which gives us $N_b = (4/3 \pi A_0^3 \rho)^{1/(1-3\nu)}$. For $\rho =0.5$ $N_b \approx 3$ and for $\rho = 1$ $N_b \approx 1$. Since for monomer densities used in our simulations the average number of monomers per blob is between $1$ and $3$, it is obvious that the blob-scaling picture cannot be valid for these densities, and monomer-monomer interactions determine the excess energy due to confinement.

In SRD the polymer is immersed in a solvent. Accordingly, the excess energy due to confinement should be derived starting from the free energy of the mixing of the polymer and the solvent. We may approximate this by the mixing free energy per lattice site for a polymer in Flory-Huggins theory $f_m(\phi) = N^{-1} \phi \ln(\phi) + (1-\phi)\ln(1-\phi) + \chi (1-\phi)$, where $\chi$ is a constant~\cite{doi}. Flory-Huggins theory calculates this energy for a polymer consisting of hard spheres. Here, $\phi$ is the fraction of lattice sites occupied by polymer segments. $f_m$ diverges as $(1-\phi) \ln(1-\phi)$, when $\phi \to 1$. For the hard-sphere lattice polymer the prefactor would go to zero as $\phi$ approaches $1$. For our continuum model, where soft potentials are used for the polymer the prefactor for the effective $\phi \propto \rho$  is not exactly of this form and does not have to vanish for $\phi_{eff} =1$. Fig.~\ref{fig:forceVsDensity}~(a) shows the divergent term of the Flory-Huggins mixing energy and the exponential form fitted to the measured force at the pore $f$. $f = C \exp[B N(t)]$ gives the best fit, but also the Flory-Huggins form aligns with the data reasonably well. We conclude that the mixing entropy explains our measured free energy due to confinement. The deviation from the exact form derived exactly within the Flory-Huggins theory arises due to the soft interaction potentials and the continuous space in our model.~\cite{takahiro}

\subsection{Time dependence of the monomer number in the capsid}

Using the measured dependence $f = C \exp[A N(t)/N_0]$, we can derive $N(t)$ in the framework presented in~\cite{sakaue_polymer_decompression}. If the monomers are packed inside the capsid by force applied at the pore, then the excess energy that results from packing $N$ monomers is $\Delta F \approx \sum_{i=0}^{N-1}  f_i \Delta l_i$, where $f_i$ is the force required to move the bead $i$ at the inner pore opening into the capsid and the bead $i+1$ in its place. $\Delta l_i$ is the distance the bead $i+1$ needs to be moved. Since $f_i$ is measured for individual beads, $\Delta l_i = a \approx 1$, $\forall i$. In the limit $\Delta l_i \to 0$ and the minimum necessary force being applied continuously on the polymer at the inner pore opening  $\Delta F = \int_1^N f(n)dn = \int_1^NC \exp(A n/N_0) dn = (C N_0/A) [\exp(BN) - D]$, where $A$, $C \propto k_BT$, and $D$ are constants. Relating the rate of change of this energy to the overall dissipation (see text after Eq.~(\ref{conf_en})), $\Delta \dot{F}(t) = -T\dot{S}(t) \approx \eta[a\dot{N}(t)/\xi(t)]^2\xi(t)^3$, we get
\begin{equation}
\dot{N}(t) = -\frac{C}{\eta a^2 \xi(t)} e^{AN(t)/N_0}.
\label{Nrate}
\end{equation}  

For the high monomer densities monomers close to the pore are pushed out from the dense initial conformation. Hence for the large part of the ejection the correlation length is not expected to change appreciably. Approximating the correlation length to be constant $\xi(t) = \xi$ and using the initial condition $N(t=0) = N_0$, the solution is given in the form
\begin{equation}
N(t) = -\frac{N_0}{A} \ln{\Big(\frac{AC}{N_0\eta a^2 \xi}t + e^{-A}\Big)}.
\label{Ntime}
\end{equation}

Fig.~\ref{fig:forceVsDensity}~(b) shows the fit of Eq.~(\ref{Ntime}) to the measured $N(t)$. A fit of $N(t)$ given by Eq.~(\ref{sakaue_nt}) for $N_0 = 283$  is given for reference. Eq.~(\ref{Ntime}) is seen to describe $N(t)$ with excellent precision as it should given that $t_w \propto 1/f$ almost throughout the ejection.

\subsection{Ejection time grows linearly with polymer length}

For constant $\rho_0$, the waiting times $t_w$ as functions of the normalized coordinate $s/N_0$ fall on the same curve except for the final retraction, see Fig.~\ref{fig:waitingTime}~(d), so for this part $\tau \sim N_0$. The following calculation also confirms this: Figs.~\ref{fig:waitingTime}~(a) and (b) show the times $\tau_2^*$ when the measured force exerted by the polymer segment inside the capsid on a monomer at the pore is zero. These points, determined by measuring the force at the pore needed to keep segments of different lengths $N$ completely inside the capsid, obey $N(t=\tau_2^*) = KN_0$, where $K \approx 0.38$. Solving Eq.~(\ref{Ntime}) for $t = \tau_2^*$ gives

\begin{equation}
\tau_2^* = \frac{N_0 \eta a^2 \xi}{AC} (e^{-KA} - e^{-A}) = P N_0,\ P > 0
\label{startime}
\end{equation}
for the pressure-driven part. This linear dependence does not result from $N(t=\tau_2^*) \sim N_0$ but is mainly due to $\Delta F$ not scaling with $N$, which can be shown by replacing $\Delta F$ of Eq.~(\ref{conf_en}), by our measured exponential form in the framework presented in~\cite{sakaue_polymer_decompression}. This leads to linear dependence for large $N_0$.

From Fig.~\ref{fig:waitingTime}~(d) it is seen that $t_w(s/N_0)$ for different $N_0$ deviate from the common form only at the final stage of ejection when the remaining part of the polymer retracts from the capsid. Retraction speed increases identically for all polymers. After $\tau_2^*$ the force at the pore due to entropic imbalance between the outside and inside of the capsid causes the tension to propagate from the pore along the polymer segment in the capsid. The tensed segment $\Delta s$ grows identically for all $N_0$. Hence, $\Delta s/N_0$ will be larger for smaller $N_0$ resulting in retraction starting at smaller $s/N_0$ for shorter polymers, as seen in Fig.~\ref{fig:waitingTime}~(d). This finite-size effect results in superlinear scaling $\tau \sim N_0^\beta$, $\beta>1$.  Accordingly, for asymptotically long polymers linear dependence $\tau \sim N_0$ would be obtained.

\subsection{Final remarks}

Finally, it is in place to note that to determine the free energy of the polymer in a very strong spherical confinement, as we have done, it is essential to use the generic FJC polymer model. Else, for example the validity of the blob-scaling picture could not be evaluated. FJC is the relevant model for example when simulating the ejection of ssDNA or RNA. In spite of our generic polymer model the situation is in many respects analogous to a realistic dsDNA in a capsid. The packing force for the dsDNA in a capsid is mainly determined by the long-ranged electrostatic interactions. Hence, in both our model and the dsDNA the dominating interactions are not the interactions of blobs but interactions of individual monomers via repulsive potentials, in our case the Lennard-Jones potential and in the case of a DNA the Coulombic potential. Exponential-looking dependencies of the packing force on monomer concentration have, indeed, been obtained experimentally, see {\it e.g.}~\cite{smith}. Also in the related computational study a dependence resembling exponential was found~\cite{kindt_dna_packaging_forces}. However, since the dependencies were plotted in linear scale and the ranges were quite small, dependency in these two studies cannot be determined with certainty. A recent computational study for semiflexible polymers also shows a packaging force dependence that does not scale with $N_0$~\cite{polson_ellipsoidal_cavity}.

\section{Conclusions}

In summary, we have investigated in detail the ejection of flexible polymers from spherical capsids through a nanoscale pore via computer simulations using realistic dynamics. The ejection dynamics and the pertaining excess energy due to confinement $\Delta F$ were analyzed via measured waiting time profiles and forces exerted on polymers at the pore. We found that the waiting times $t_w$ grow exponentially with the number of ejected monomers and that the force $f$ measured at the pore increases exponentially with the number of monomers in the capsid $N$. We showed that $\Delta F$ then must grow exponentially with $N$, which we addressed to be due to the higher-order terms in monomer-monomer interactions. This exponential dependence is very similar to the divergence of the mixing energy in the Flory-Huggins theory at high concentration. We also found that $t_w \sim 1/f$ holds well for the simulated densities and that, accordingly, $\Delta F$ determines the ejection dynamics for such strongly confined polymers. We showed that this $\Delta F$ results in the ejection time $\tau$ growing linearly with the polymer length $N_0$. The measured superlinear scaling $\tau \sim N_0^\beta$, $\beta > 1$, results from a finite-size effect due to the final retraction of polymers' tails from capsids.

Importantly, for densities that are larger than those associated with strong confinement but still moderate compared with realistic densities the strong monomer-monomer interactions result in the excess energy due to confinement increasing exponentially with the number of monomers in the capsid and ejection time increasing linearly with polymer length. This is in stark contrast with the previous results on ejection dynamics that are valid only in the semidilute regime where blob-scaling picture applies. Also of importance is that, contrary to some claims, for these densities the ejection easily completes without any assisting mechanism such as flow.

\begin{acknowledgments}
We thank T. Sakaue for useful comments. The computational resources of CSC-IT Centre for Science, Finland, and Aalto Science-IT project are acknowledged. The work of Joonas Piili is supported by Tekniikan edist\"amiss\"a\"ati\"o and The Emil Aaltonen Foundation.
\end{acknowledgments}

\bibliography{references.bib}

\end{document}